# Plasma asymmetry, electron and ion energy distribution function in capacitive discharges excited by tailored waveforms

Sarveshwar Sharma[1,2,*], Nishant Sirse[3,*], Animesh Kuley[4] and Miles M Turner[5]

[1]Basic Theory and Simulation Division, Institute for Plasma Research, Gandhinagar – 382428, India

[2]Homi Bhabha National Institute, Anushaktinagar, Mumbai – 400094, India

[3]Institute of Science & Research, IPS Academy, Indore -452012, India

[4]Department of Physics, Indian Institute of Science, Bangalore, Karnataka, -560012, India

[5]School of Physical Sciences and National Center for Plasma Science and Technology, Dublin City University, Dublin 9, Ireland

[*]E-mail: nishantsirse@ipsacademy.org , sarvesh@ipr.res.in

**Abstract**

Using particle-in-cell simulation technique, we investigate the plasma and ionization asymmetry, electron and ion energy distribution function in capacitive discharges excited by tailored waveforms. At a base frequency of 13.56 MHz, three different waveforms namely, sinusoidal, saw-tooth, and square are applied for a constant current density of 50 A/m2 and 5 mTorr argon gas pressure. The simulation results show that the square waveform produces the highest plasma density in the discharge, whereas maximum asymmetry is observed for plasma excited by sawtooth like waveform. Both square and sawtooth waveforms generate multiple beams of high-energy electrons from near to the expanding phase of the sheath edge followed by the high-frequency modulations up to 100 MHz on the instantaneous sheath position. The electron energy distribution function depicts 3 electron temperature and highly elevated tail-end electrons for the square waveform in comparison to the sinusoidal and sawtooth waveform. The ion energy distribution function is bimodal at both powered and grounded electrodes with a large asymmetry and narrow type distribution in the case of sawtooth like waveform. These results suggest that the choice of the waveform is highly critical for achieving maximum asymmetry and plasma density simultaneously in the discharge.

1. Introduction

Low pressure capacitively coupled plasma (CCP) discharges can be operated by either current or voltage driven radio-frequency (RF) waveforms applied between pair of electrodes. In semiconductor industries, CCP is one of the most important plasma processing tools for the fabrication of large-scale integrated circuits [1]. In order to achieve uniform processing over the substrate, it is crucial to have control on some of the critical plasma sheath parameters like the ion flux and ion energy. Higher ion flux is required for enhancing the processing rate that depends on the plasma density, whereas lower energy is beneficial for preventing surface damages. In CCP discharges, operated by single frequency sinusoidal waveforms the higher densities can be achieved either by applying higher voltages or higher frequencies by keeping all other controlling parameters constant [1,2,3,4,5,6,7,8,9,10,11,12,13,14,15, 16]. On the other hand, a Dual-frequency waveform operated CCP discharge produces high-density plasmas either by the variation of high frequency voltage amplitude or the higher frequency itself also provides an additional control on the ion energy by changing the lower frequency parameters [17,18,19, 20,21,22,23,24,25,26,27,28,29, 30].
  Generation of plasma using non-sinusoidal waveforms is an another alternative for achieving control on the plasma sheath parameters and asymmetry in the discharge [31,32,33,34,35,36,37,38, 39, 40, 41, 42]. A popular terminology used for these non-sinusoidal waveforms are 'tailored waveforms' that can be created by the superimposition of fundamental frequency and its higher harmonics with suitable phase shift between them [43, 44, 45, 46]. The benefits of tailored waveform excited CCP discharges were first demonstrated by Patterson et al [31] where he showed that by increasing the number of harmonics the plasma density and ion flux increases significantly. Using Hybrid simulation, Heil et al [32, 33] showed that a DC self-bias can be generated even in in a geometrically symmetric reactor using a fundamental frequency and its second harmonic, called as electric asymmetry effect (EAE). Donko et al further demonstrated that using phase separated waveforms the ion flux stays nearly constant and the self-bias, and hence the maximum ion energy changes significantly [34]. Further studies performed by increasing the number of harmonics in the tailored waveform has shown that the ion flux and average ion energy on one of the electrodes can be enhanced without changing it on to the other electrode [47,48,49,50]. This effect was attributed to the ionization asymmetry that produce more intense ionization on one of the electrodes as compared to the other electrode. In processing applications, the use of tailored waveforms played an important role in thin film deposition [51, 52] and silicon etching [53, 54].

  The use of temporal asymmetric waveform for generating an asymmetric plasma response was also investigated. In recent years, several simulation and experimental studies have been performed to investigate the CCP discharges excited by temporal asymmetric waveform, particularly using a saw-tooth like waveform [38, 39, 40, 42]. A simulation study performed by Bruneau et al [38] used different rise and fall slopes of sawtooth voltage waveforms and demonstrated that high ionization rate occurs in the vicinity of sheath edge near to the powered electrode because of the rapid sheath expansion. The ionization spatial asymmetry obtained using such waveforms has been demonstrated experimentally using phase resolved optical emission spectroscopy (PROES) [40]. Further studies performed at different gas pressure showed that the flux asymmetry decreases at low gas pressure [39]. Using 1D3V PIC method, Sharma et al. [42] demonstrated the influence of driving frequency on plasma density, discharge asymmetry and electric field nonlinearity for sawtooth like current waveform. It was shown that the formation of high frequency sheath edge oscillations at lower RF driving frequency creates higher plasma density. The burst like structures in time averaged J.E i.e., alternate electron heating and cooling) was observed, which suggested that the heating mechanism cannot be described by using simple analytical model [42]. One of the studies performed by Donko et al compared ion energy and angular distribution in CCP excited by single-frequency, classical dual-frequency, valleys and sawtooth-type waveforms. Their results shows that the valley-type waveform produces controlled narrow ion energy distribution when compared to the classical dual-frequency waveform. On the other hand, sawtooth-type waveform has shown minimal effect on the angular distributions.

Most of the previous works using tailored focuses on one kind of waveform, however, there exist very few studies comparing different tailored waveforms and its comparison with sinusoidal waveform. In the present work, we investigate the discharge asymmetry, electron beam generation and high frequency sheath modulation along with the electron and ion energy distribution function in CCP discharge using 2 different temporally asymmetric waveforms namely sawtooth-type and square waveform, and its comparison with plasma excited by sinusoidal waveform at a fundamental RF frequency of 13.56 MHz.

The paper is organized as follows. The details of simulation parameters and description of simulation technique that is based on Particle-in-Cell/Monte Carlo collision (PIC/MCC) methods is given in section II. The physical explanation of simulation results is described in section III. The summary and conclusion of paper is given in section IV.

## 2. Simulation Technique and Parameters

A self-consistent 1D3V electrostatic particle-in-cell (PIC) code has been used to simulate a symmetric CCP discharge for the argon plasma. This code is well tested, benchmarked and developed by Prof. Miles Turner at Dublin City University. It is comprehensively used in several research papers and articles and few of the important references can be find here [11,14,55, 56, 57, 58, 59, 60, 61, 62, 63]. The code is based on the Particle-in-Cell/Monte Carlo collision (PIC/MCC) technique and essential features of this method are given in the literature [64, 65]. This code can handle both current and voltage driven cases and for the present study we have used the *current driven* mode. The particulars of the simulation procedure adopted in the code are reported in the Ref. [66]. All the critical particle-particle collision reactions *viz.* ion-neutral (inelastic, elastic and charge exchange), electron-neutral (ionization, elastic and inelastic) and processes like metastable pooling, multi-step ionization, super elastic collisions, partial de-excitation and further de-excitation. The different type of reactions included in all set of simulations is given in Ref. [61]. So, the production of metastables (Ar$^*$ and Ar$^{**}$) are considered and tracked in the output diagnostics. In simulation we have considered the charge particles like electrons, ions and the two lumped excited states of Ar (Ar$^*$ (3p$^5$4s), 11.6 eV, and Ar$^{**}$ (3p$^5$4p), 13.1 eV) with neutral argon gas background. The particulars of the species considered in simulation, the detailed reactions and the cross sections are taken from well-tested sources and can be find in the literature [57,61, 67]. The grid size (Δx) and time step size (Δt) is chosen in such a way that can resolve the Debye length ($\lambda_{De} = \sqrt{\varepsilon_0 k_B T_e / n_e e^2}$) and the electron plasma frequency ($f_{pe} = \sqrt{n_e e^2 / \varepsilon_0 m_e}/2\pi$) respectively so that the stability and accuracy criterion of PIC can be satisfied. The electrodes considered here are planar and parallel to each other having infinite dimension. The electrodes are assumed perfectly absorbing and secondary electron emission is also ignored for the sake of simplicity. The gap between electrodes is 6 cm and the operating neutral gas pressure is 5 mTorr. In simulation, the neutral argon gas is uniformly distributed having fixed temperature 300 K. The ion temperature is also same as neutral gas temperature. The number of particles per cell for all set of simulations is 100. All the simulations run for more than 5000 RF cycles to achieve the steady state profile.

Three different types of current waveforms namely sinusoidal, sawtooth and square used here. The complex structures of waveforms like sawtooth and square can be generated by superimposition of a reasonable number of harmonics of a fundamental RF frequency and also logically picked the phase shifts between them. For current driven CCP, following are the mathematical expression for sinusoidal, sawtooth and square waveforms (see figure 1):

$$J_{rf}(t) = J_0 \sin(\omega_{rf} t) \quad \text{----------} \quad (1)$$

$$J_{rf}(t) = \pm J_0 \sum_{k=1}^{N} \frac{1}{k} \sin(k\omega_{rf} t) \quad \text{----------} (2)$$

$$J_{rf}(t) = J_0 \sum_{k=0}^{N} \frac{\sin[(2k+1)\omega_{rf} t]}{2k+1} \quad \text{----------} (3)$$

where in eq. (2) the positive and negative signs represent to "sawtooth-down" and "sawtooth-up" waveforms respectively. Here, $J_o$ is the amplitude of current density, which is applied at the powered electrode, and $\omega_{rf}$ is the fundamental angular driven radiofrequency. The magnitude $J_0$ varies with the total number of harmonics N (which is 50 here for sawtooth and square waveforms) and is organized in a way to construct the required peak-to-peak current density amplitude. We have used "sawtooth-down" current waveform. Figure 1 shows the profile of sinusoidal, sawtooth and square waveforms for base frequency 13.56 MHz with current density amplitude of 50 A/m$^2$. The choice of current driven CCP discharge is arbitrary here because the main motivation of present research work is to investigate the effect of different current waveforms (sinusoidal, sawtooth and square) on the plasma parameters and sheath dynamics. Moreover, most of the analytical models are based on current driven cases so direct comparison of simulation results with them is relatively straightforward [68, 69].

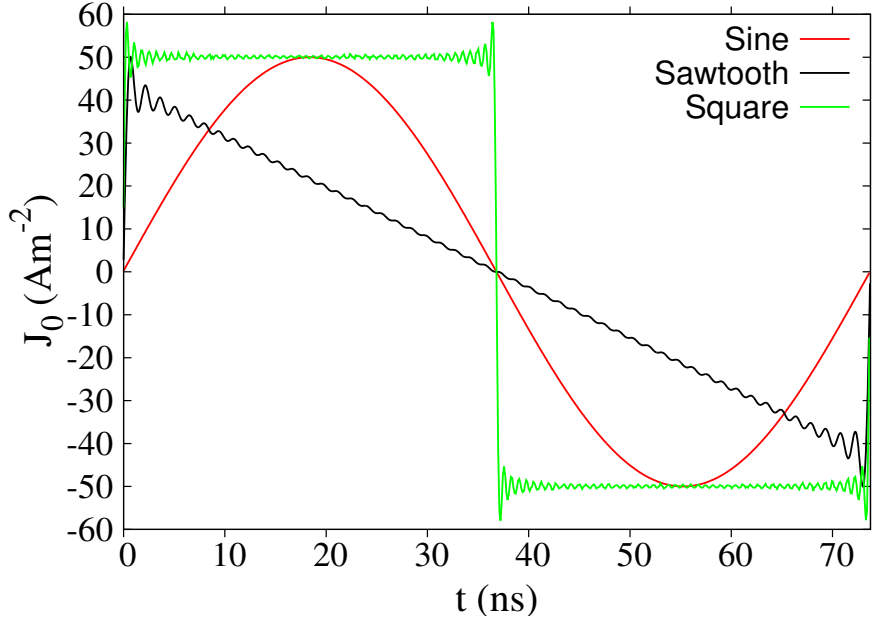

**Figure 1:** Current profile of sinusoidal, sawtooth and square waveforms for base frequency 13.56 MHz having current density amplitude of 50 A/m².

## 3. Results and Discussions

For a constant RF driving frequency, the shape of applied waveform greatly alters the sheath dynamics and therefore the plasma profile eventually modifies. Figure 2 (a) presents the spatial profile of time averaged electron and ion density in the discharge for different current waveforms i.e., sinusoidal, sawtooth, and square. The fundamental RF driving frequency is 13.56 MHz and the current density amplitude is 50 A/m² for all the cases. The argon pressure is kept constant at 5 mTorr for all set of simulations. In figure 2 (a), the RF current source is applied at L = 0 mm and the electrode at 60 mm is grounded. As shown in figure 2 (a), the central plasma density is ~8.4×$10^{15}$ $m^{-3}$, ~1×$10^{16}$ $m^{-3}$, and ~2.1×$10^{16}$ $m^{-3}$ for sinusoidal, sawtooth, and square waveform respectively. Thus, the density formation is highest in the case of square waveform, which is typically 2.5 times higher compared to the sinusoidal waveform. Corresponding ion flux measured at the grounded electrode is 7.5×$10^{18}$ $m^{-2}s^{-1}$ and 1.7×$10^{19}$ $m^{-2}s^{-1}$ for sinusoidal and square waveform respectively which indicates that the latter is ~2.3 times more compared to the earlier. As discussed in section I, higher plasma density and therefore an enhanced ion flux at the electrode is the fundamental criterion for an improved plasma processing rate. Therefore, the square waveform is far better for achieving the higher processing rate compared to traditional sinusoidal waveform. In contrast to the higher plasma density observed in the case of square waveform, the sawtooth waveform presents larger bulk plasma length. The sheath width at the powered/grounded electrode are 6 mm/9.7 mm and 10.4 mm/9.5 mm for sawtooth and square waveforms respectively. For sinusoidal waveform, the sheath width near to the electrodes are approximately the same (~15 mm). The sheath width is calculated by observing the maximum position of the electron sheath edge from the electrode and where the quasi-neutrality breaks down. Coming to the plasma asymmetry it is observed that the sawtooth waveform shows maximum asymmetry. The corresponding DC self-bias is highest for the sawtooth waveform (~210 V) that represents the strong asymmetry, whereas the square waveform generates minimal negative DC self-bias of ~50 V. On the other hand, the DC self-bias is negligible in the case of sinusoidal waveform and therefore no discharge asymmetry is present. A plot of time-averaged potential profile for all the waveforms is displayed in figure 2 (b) showing the formation of DC self-bias on the powered electrode.

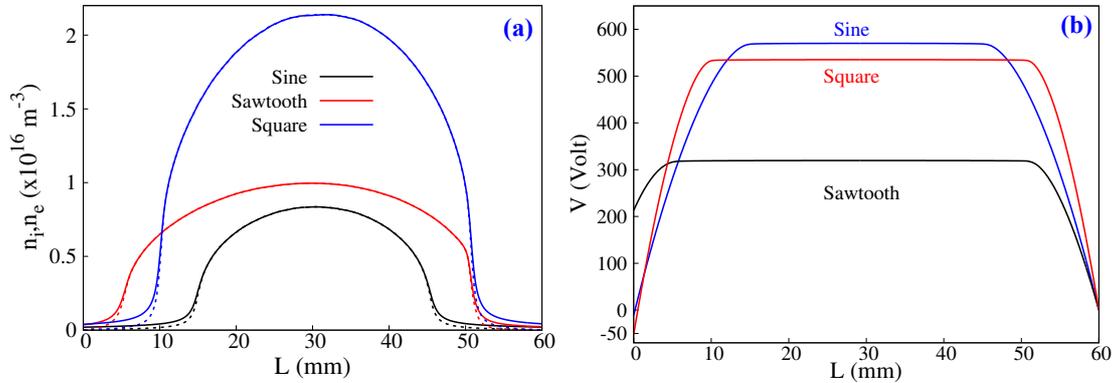

**Figure 2:** (a) The spatial profile of time average ion (solid line) and electron density (dashed line) and (b) time-averaged potential profile for different waveforms (sine, sawtooth and square) at 13.56 MHz applied frequency. The current density amplitude is 50 A/m$^2$.

The plasma density enhancement by changing the profile of current waveform can be understood by analyzing the ionization rate inside the discharge system. Figure 3 shows the time averaged ionization rate along with the excitation rates at current density amplitude of 50 A/m$^2$ for three different waveforms. At a low gas pressure, the electron mean free path is longer and hence the ionization and excitation mechanisms are mainly appearing inside the bulk plasma. Figure 3 (a) shows the direct ionization (e + Ar → 2e + Ar$^+$) while electron collides with ground state argon atom. It is clear from figure that the overall direct ionization rate is highest for the square waveform. The sinusoidal waveform has higher ionization rate in the bulk plasma when compared to sawtooth waveform, however, it is drastically lower near to the electrodes i.e., up to 15 mm from both electrodes. Figure 3 (b) and (c) shows the multi-step ionization occur when electron impact with metastable state Ar$^*$ (e + Ar$^*$ → e + Ar$^+$) and Ar$^{**}$ (e + Ar$^{**}$ → e + Ar$^+$) respectively. Although contribution from such processes is lower in comparison to the direct ionization (figure 3 (a)), we can observe that the multi-step ionization process rate is highest in the case of square waveform. Comparing sawtooth with sinusoidal, we can distinctly observe that the sawtooth waveform has highest ionization rate for both cases here. These two processes along with the direct ionization create higher plasma density in case of sawtooth waveform compared to the sinusoidal profile. Figure 3 (d) and (e) shows the production of metastable Ar$^*$ (e + Ar → e + Ar$^*$) and Ar$^{**}$ (e + Ar → e + Ar$^{**}$) from neutral gas. The square waveform has a very high population of Ar$^*$ ($3p^54s$, 11.6 eV) and Ar$^{**}$ ($3p^54p$, 13.1 eV) in all three cases. Although the central metastable densities (Ar$^*$ and Ar$^{**}$) is higher for sinusoidal in comparison to the sawtooth waveform, however the average metastable densities production (Ar$^*$ 5.5×10$^{19}$ $m^{-3}$ Sine 6.5×10$^{19}$ $m^{-3}$ sawtooth Ar$^{**}$ 1.9×10$^{20}$ $m^{-3}$ sine 2.3×10$^{20}$ $m^{-3}$ sawtooth) is higher for sawtooth waveform. It is also important to notice from figure 3 (d) – 3 (f) that the production of Ar$^{**}$ is higher compared to Ar$^*$ in all the cases. The low energy electrons (~1.5 eV) are further absorbed in excitation from Ar$^*$ to Ar$^{**}$ (e + Ar$^*$ → e + Ar$^{**}$) and its production rate is utmost in square waveform and lowest for sinusoidal waveform (see figure 3 (f)).

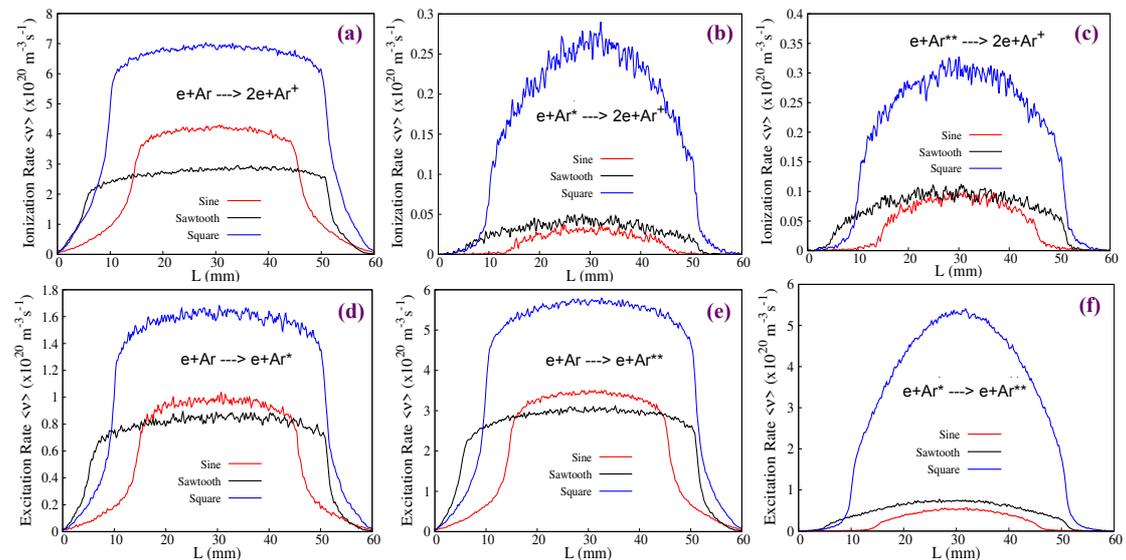

**Figure 3:** Time averaged ionization ((a)-(c)) and excitation ((d)-(f)) rates at different applied current waveforms (sinusoidal, sawtooth and square) for 13.56 MHz RF frequency and current densities of 50 A/m$^2$.

The observed asymmetries when applied waveform is changed is mainly due to an asymmetry in the spatio-temporal ionization dynamics. To elucidate this, we further investigate the ionization mechanism by examining the spatio-temporal profile of ionizing collision rate for three different current waveforms. This is plotted in figure 4 for 2 RF periods averaged over last 600 RF cycles. As shown in figure 4 (a), for sinusoidal waveform, the ionization profile is symmetric, and the maximum ionizing collision rate is observed during the time of sheath expansion at both grounded and powered electrodes. Moreover, the ionizing front penetrates the bulk plasma and reaching to the opposite sheath because of the nearly collisionless operating conditions. Figure 4 (b) shows the spatio-temporal profile of ionizing collision rate for sawtooth waveform, which has a highest asymmetry (also see figure 1) i.e., a strong ionization is observed near to the grounded sheath when compared to the power electrode sheath. Here, multiple beams of energetic electrons emerged during the expanding phase of sheath near to the grounded electrode that penetrates the bulk plasma and approach to the opposite sheath edge. The spatio-temporal profile of ionizing collision rate for square waveform is displayed in Figure 4 (c). Here the profile turn into nearly symmetric, and multiple bursts of high energy electrons emerges from the expanding phase of the sheath edge. The intensity of these bursts is higher in comparison to the sawtooth and sinusoidal waveforms. Comparing sheath velocities on the grounded electrodes for different waveforms shows it is highest for square waveform ($7.35 \times 10^5 m/s$) and lowest for sinusoidal waveform ($2.2 \times 10^5 m/s$). The sheath velocity for sawtooth waveform is $4.97 \times 10^5 m/s$. The higher sheath velocity enhances the energy gained by the oscillating electrons and henceforth ionizing collision rate and therefore the plasma density also goes up. Further by looking at the figure of spatio-temporal ionizing collision rate, it is clearly observed in figure 4 (b) that multiple beams like structures are present near to the grounded electrode during the sheath expansion. Similarly, in figure 4 (c) intense multiple beams like structures are created both near to the grounded and powered electrode during the time of sheath expansion. However, such multiple beams like structures are absent in the case of sinusoidal waveform. The reason for the creation of these multiple beam structures during the sheath expansion is due to the high frequency modulation on the instantaneous sheath edge position which is similar to the sheath modulation in CCP discharges excited by multiple frequencies [26, 28, 70]. We will support this argument by investigating the non-linearity in the spatio-temporal profile of electric field and higher harmonics generation on the instantaneous sheath edge position.

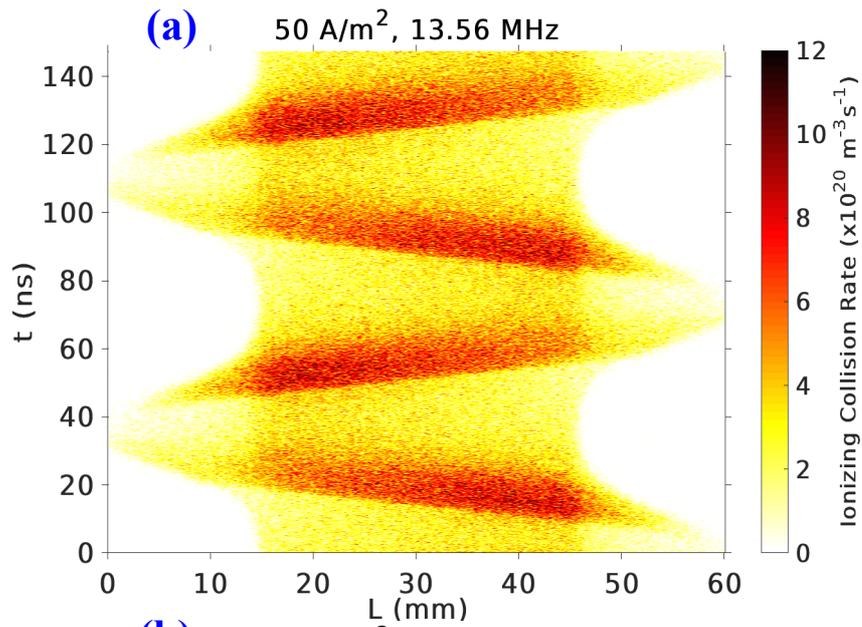
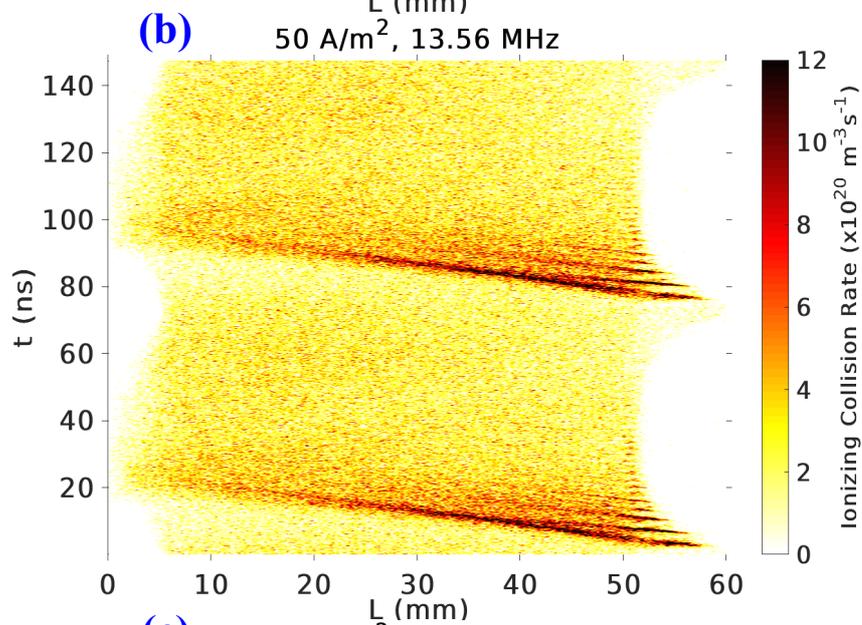
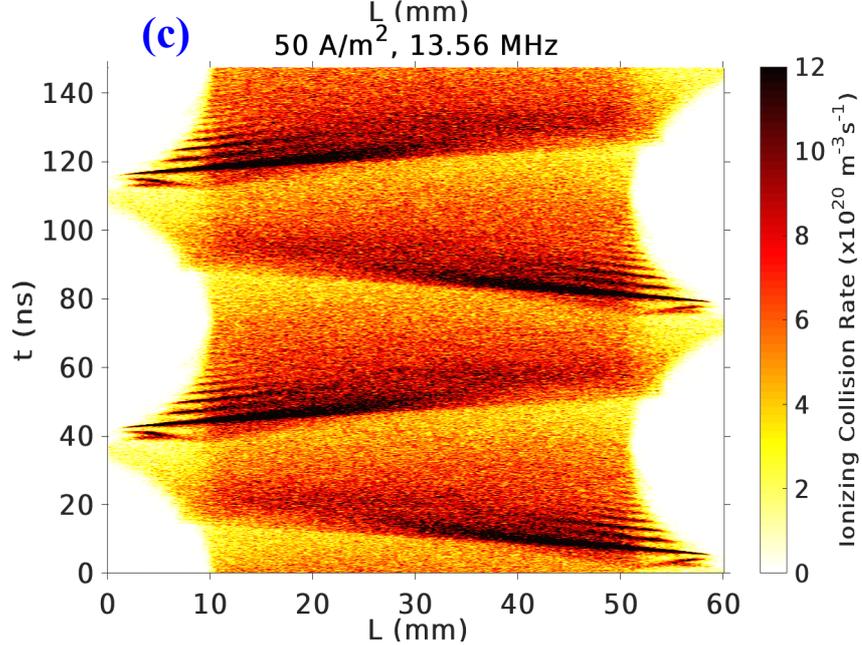

**Figure 4:** Spatio-temporal evolution of ionizing collision rates at different current waveforms i.e. (a) sinusoidal (b) sawtooth and (c) square for 13.56 MHz for current densities of 50 A/m$^2$.

Figure 5 (a), 5 (b) and 5 (c) shows the spatio-temporal evolution of electric field for the sinusoidal, sawtooth and square waveforms respectively. The data shown here is for 2 RF periods, averaged over last 100 RF cycles when the simulation results achieved steady state condition. As shown in figure 5 (a), for sinusoidal case, the electric field is mostly confined in the sheath region and the bulk plasma is almost quasi-neutral. The instantaneous sheath edge position is smooth in this case. When the applied current waveform is changed to sawtooth (see figure 5 (b)) waveform, the temporal electric field at the sheath boundary near to the grounded electrode is substantially modified i.e., high frequency oscillations are observed on the instantaneous sheath edge position. The electric field here is also confined inside the sheath region and bulk plasma is almost quasi-neutral. Furthermore, when the applied current waveform is square (see figure 5 (c)), the temporal electric field at both the sheath edges, i.e., near to the grounded and powered electrode, are significantly modified and the high frequency oscillations are clearly visible on the instantaneous sheath edge positions. These high frequency modulations on the instantaneous sheath edge at both the sheath edges is one of the principal factors responsible for driving the higher ionization rate and as a result producing higher plasma density in the system. The presence of electric field inside bulk plasma is also significant here which is termed as "electric field transients". The formation of these electric field transients is reported in earlier articles [11,14,42,71] and mainly due to the higher sheath velocities that enhances energy gained by the electrons interacting with the oscillating sheath. Such electric field transients were mostly observed at higher driving frequencies [14,62], whereas here it is present at a fundamental frequency of 13.56 MHz but only in the case of square waveform. We believe this is mostly related with the sheath velocity and higher harmonics on the instantaneous sheath position, which is higher in the case of square waveform and discussed in the following paragraph.

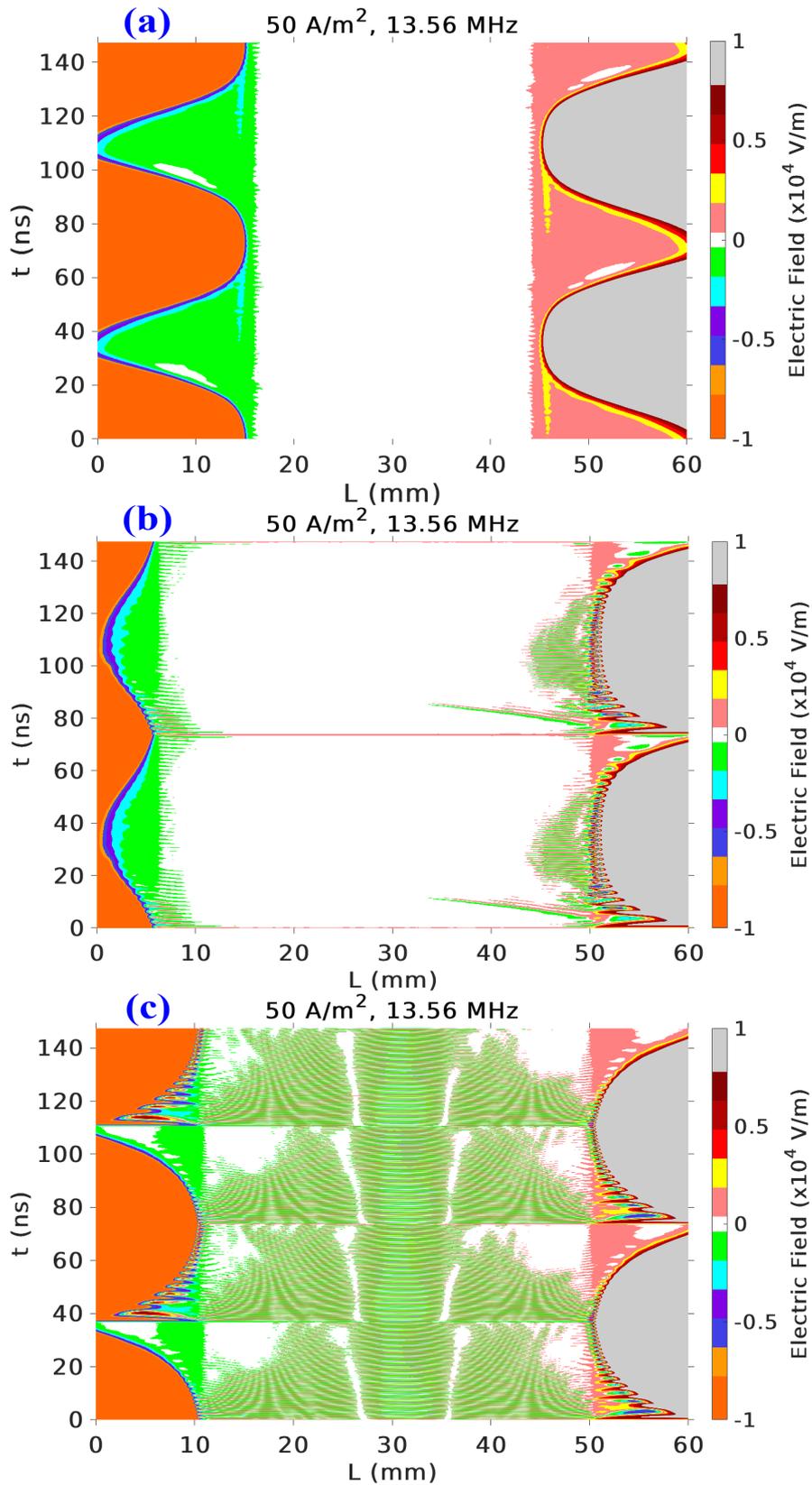

**Figure 5:** Spatio-temporal evolution of electric field for 2 RF cycles at different applied current waveforms for 50 A/m$^2$ and 13.56 MHz: (a) Sinusoidal (b) Sawtooth (c) Square.

Figure 6 (a) and 6 (b) showing the instantaneous sheath position and its FFT respectively for different applied current waveforms. The graph 6 (a) is plotted for one RF period near to the grounded electrode. Again, the sheath width is calculated by observing the maximum position of the electron sheath edge from the electrode and where the quasi-neutrality breaks down. As shown in figure 6 (a), for sinusoidal waveform, sheath width is expanding up to a maximum position of ~15 mm from the grounded electrode. The instantaneous sheath position in this case is varying smoothly within the RF period i.e., it is reaching up to the maximum position and collapsing back towards the grounded electrode. In contrast to the sinusoidal waveform, the maximum sheath edge position is lower (~9.5-9.7 mm) in the case of sawtooth and square waveforms. Furthermore, both sawtooth and square waveforms show high frequency modulation on the instantaneous sheath position. These high frequency oscillations are similar to the multi-frequency excited CCP discharges [26,70]. Due to which the conjugate sheath velocity modifies drastically and remains higher for square waveform ($7.35 \times 10^5 m/s$) as the number of oscillations on the instantaneous sheath edge position is higher in this case. The corresponding FFT is shown in figure 6 (b). It is clear from figure 6 (b) that the contribution of fundamental frequency is highest in the case of sinusoidal waveform and higher harmonics contribution drops monotonically. Whereas higher harmonic contents up to the 7$^{th}$ harmonics are clearly visible for sawtooth and square waveforms (figure 6 (b)). While comparing different waveforms, it is observed that the higher harmonic content (above 3$^{rd}$ harmonics) is maximum in the case of square waveform. Thus, the presence of higher harmonic contents justifies production of higher plasma density in the case of square waveform as they are efficient in the power deposition. It is noteworthy that due to asymmetry, the sawtooth waveform posses high frequency oscillations during the initial phase of the collapsing sheath edge (0.5-0.8 on Y-axis in the RF period) and then smoothen out towards the grounded electrode.

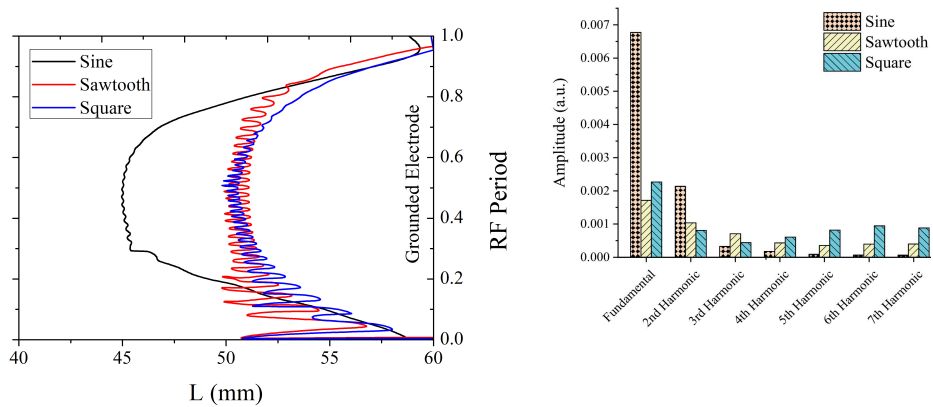

**Figure 6:** (a) Instantaneous sheath edge position for one RF period and (b) FFT of instantaneous sheath edge for different applied current waveforms (sinusoidal, sawtooth, and square) for 13.56 MHz RF frequency and current densities of 50 A/m$^2$.

The presence of higher harmonics on the instantaneous sheath edge position and therefore the higher sheath velocity drives the higher electron energy beams from near to the sheath edge. As discussed earlier the sheath velocity is lowest for sinusoidal waveform ($2.2 \times 10^5 m/s$) and highest for square waveform ($7.35 \times 10^5 m/s$). These high-energy electrons create the electric field transients (figure 5 (c)) inside bulk plasma, which increase the population of high energy electrons through non-linear interaction. This is cleared by observing the *electron energy distribution function* (EEDF) at the center of discharge. Figure 7 shows the EEDF for 3 different waveforms. As displayed in figure 7, both the low energy electrons and high energy tail-end electrons are highest in the case of square waveform. The shape of the EEDF for square waveform reveals 3 electron temperatures i.e., bulk electron temperature (~ 6 eV), mid energy range (~ 6 eV to 33 eV) and high energy tail (> 33 eV). The sawtooth waveform produces higher population of low energy electrons when compared to sinusoidal waveform, however, its tail-end electron population is slightly lower than sinusoidal waveform. This effect is attributed to the asymmetric discharge behavior in the case of sawtooth waveform. As shown in the figure 5 (a) and 5 (b), the electric field profile shows that the sheath is nearly symmetric on both powered and grounded electrode for sinusoidal waveform (sheath voltage ~ 1380 V at powered and grounded electrode), whereas, for sawtooth waveform the powered electrode sheath is smaller in comparison to the grounded electrode. Thus, although the sheath velocity near to the grounded electrode is higher for sawtooth waveform and impart higher energy to the electrons, but they are not effectively confined on the opposite sheath edge as the powered electrode sheath voltage is lower (sheath voltage ~ 350 V at powered electrode and ~ 560 V at grounded electrode). On the other hand, the total ionization rate is also higher for sawtooth waveform when compared to the sinusoidal waveform. The combined effect of these two mechanisms produces lower tail-end electron population for sawtooth waveform.

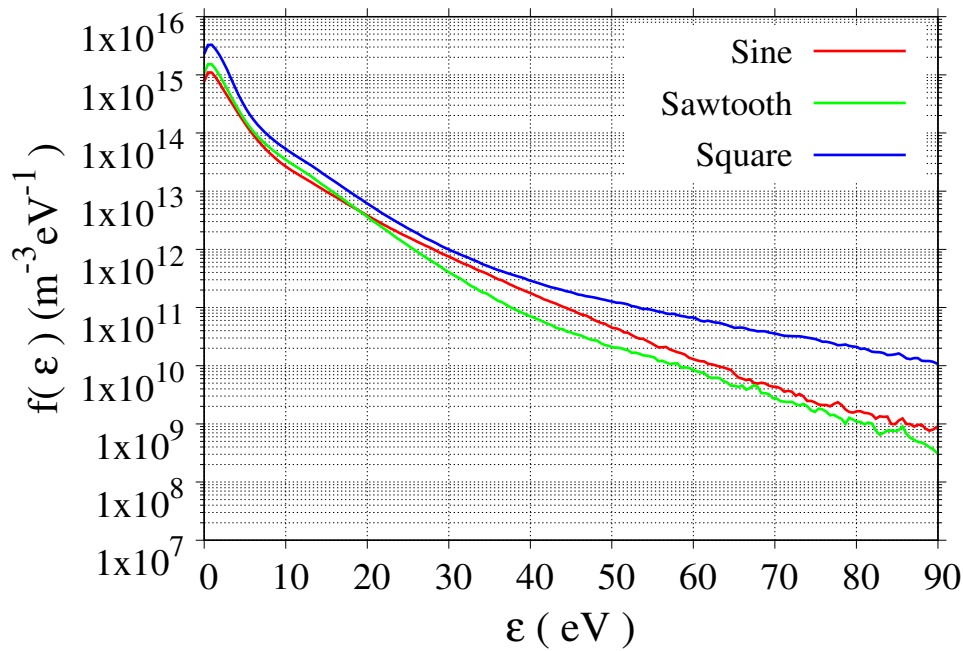

**Figure 7:** Electron energy distribution function for three different current waveforms at the center of discharge. The current density amplitude is 50 A/m$^2$ and the RF frequency is 13.56 MHz.

Next, we study the effect of different current waveforms on the ion energy distribution function (IEDF) at the electrode surface. Figure 8 (a) and 8 (b) shows the IEDF for sinusoidal, sawtooth, and square waveform at the powered and grounded electrode respectively. The shape of the IEDF is bi-modal for all the waveforms. It is noteworthy that the asymmetry in the ion energy is highest in the case of sawtooth waveform i.e., the mean ion energy is lower (~106 eV) on the powered electrode when compared to the grounded electrode (~317 eV). This asymmetry is attributed to the formation of DC self-bias and therefore the mean ion energy corresponding to the plasma potential (~320 V) at the grounded electrode and plasma potential with respect to the DC self-bias (~213 V) at the powered electrode. On the other hand, both sinusoidal and square waveform produces higher mean ion energies at both powered and grounded electrode due to nearly symmetric behavior. For sinusoidal it is nearly 570 eV, whereas for square it is approximately 539 eV at the grounded electrode and ~596 eV at the powered electrode due to the negative self-bias of ~52 V.

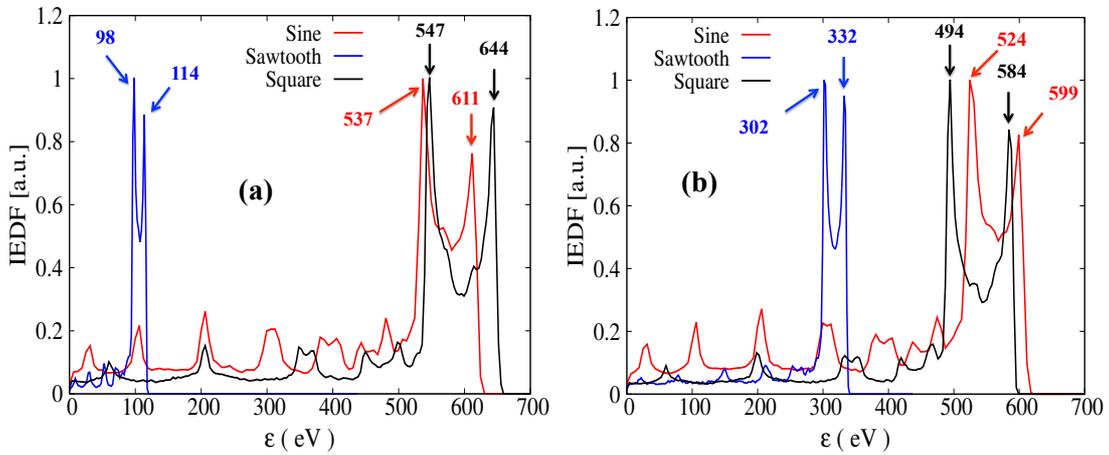

**Figure 8:** Ion energy distribution function (IEDF) for three different current waveforms at the (a) powered electrode and (b) grounded electrode. The current density amplitude is 50 A/m$^2$ and the RF frequency is 13.56 MHz

The ion energy distribution function is bimodal at both powered and grounded electrode with a large asymmetry and narrow type distribution in the case of sawtooth like waveform. These results suggests that the choice of waveform is highly critical for achieving maximum asymmetry and plasma density simultaneously in the discharge.

## 4. Summary and Conclusion

In this work we have investigated the plasma density and ionization asymmetry for sinusoidal, sawtooth and square waveforms by using particle in cell simulation technique. The base radio-frequency used here is 13.56 MHz and the CCP discharge is operated by constant current density of 50 A/m$^2$ at 5 mTorr neutral gas pressure. The simulation results demonstrates that the highest plasma density is produced by square waveform and maximum asymmetry is generated by the sawtooth like waveform. The multiple high energy electron beams are generated by the square waveform from near to the expanding sheath edge at both the electrodes. In the of sawtooth like waveform, these multiple beams are only present near to the grounded due to strong plasma asymmetry and high sheath velocity. For square waveform, the EEDF represents 3 electron temperatures with a highly elevated electron tail when compared to sinusoidal and sawtooth waveform. For sawtooth like waveform, the IEDF is bimodal at both the powered and grounded electrodes with a large asymmetry and narrow type distribution. This study propose that the choice of waveform is extremely critical for achieving maximum asymmetry and plasma density simultaneously in the CCP discharges.

**Acknowledgements**

Dr. A. Kuley is supported by the Board of Research in Nuclear Sciences (BRNS Sanctioned No. 39/14/05/2018-BRNS), Science and Engineering Research Board EMEQ program (SERB Sanctioned No. EEQ/2017/000164), National Supercomputing Mission (NSM) (Ref. No.: DST/NSM/R&D_HPC_Applications/2021/04), and Infosys Foundation Young Investigator grant.